\documentclass[twocolumn,showpacs,preprintnumbers,amsmath,amssymb]{revtex4}

\usepackage{graphicx} 
\usepackage{dcolumn} 
\usepackage{bm} 

\begin{document}


\title{Current carrying Andreev bound states in a Superconductor-Ferromagnet 
       proximity system}

\author{M. Krawiec}
\author{B. L. Gy\"orffy}
\author{J. F. Annett}
\affiliation{H. H. Wills Physics Laboratory, 
 University of Bristol, Tyndal Ave., Bristol BS8 1TL, UK}

\date{\today}

\begin{abstract}
We study the ground state properties of a ferromagnet-superconductor
heterostructure on the basis of a quasiclassical theory. We have solved the
Eilenberger equations together with Maxwell's equation fully
self-consistently and found that due to the proximity effect a 
Fulde-Ferrel-Larkin-Ovchinnikov ($FFLO$) like state is realized in such system. 
Moreover this state has oscillations of the pairing amplitude in either one or 
two directions, depending on the exchange splitting and thickness of the 
ferromagnet. In particular, using semiclassical arguments (Bohr-Sommerfeld 
quantization rule) we show that owing to the presence of the Andreev bound 
states in the ferromagnet, a spontaneous current in the ground state is 
generated as a hallmark of the $FFLO$ state in the direction parallel to the 
interface. We also discuss the effects of the the elastic disorder and finite 
transparency of the interface on the properties of the $FFLO$ state in the 
system. 
\end{abstract}

\maketitle


As is well known, ferromagnetism and singlet pairing superconductivity are 
competing phenomena. Whilst an exchange interaction favors parallel spin 
aligment, Cooper pairs must be in a singlet (spin) state. But the two phenomena 
are less mutually exclusive in artificially made ferromagnet-superconductor 
($FM$/$SC$) heterostructures \cite{Chien}. In such structures the 
ferromagnetism and the superconductivity can coexist near the $FM$/$SC$ 
interface owing to the proximity effect \cite{Lambert}. In the case when the 
normal metal is not a ferromagnet, the proximity effect has been studied for a 
long time \cite{Lambert}. By now it is rather well understood in terms of the 
Andreev reflection processes \cite{Andreev}. By contrast, the proximity effect 
in $FM$/$SC$ systems has became a centre of attention only recently. It is not 
only important from a scientific point of view, as it allows for study of the 
interplay between magnetism and superconductivity \cite{Berk}, but also from a 
technological one, as it may find applications in magnetoelectronics 
\cite{Bauer} and quantum computing \cite{Blatter}. 

A number of new phenomena has been revealed in $FM$/$SC$ multilayers. The most
interesting examples are: non-monotonic behavior of the $SC$ transition 
temperature \cite{Wong}, oscillations of a pairing amplitude 
\cite{Buzdin,Radovic,Demler} and the density of states in the $FM$ 
\cite{Kontos,Zareyan,Buzdin_1,KGA}, paramagnetic Meissner effect \cite{Lee}, 
proximity induced very long range triplet superconductivity in $FM$ 
\cite{Volkov} or generation of spontaneous currents in the ground state of such 
systems \cite{KGA}. These unusual properties, associated with Cooper pairs in 
an exchange field, can be explained in terms of an phenomenon first identified 
by Fulde, Ferrell, Larkin and Ovchinnikov ($FFLO$) \cite{FFLO}. Originally it 
has been studied in a bulk superconductor with the exchange splitting. It turns 
out that, although of great conceptual interest, the bulk $FFLO$ state can be 
realized only in a very small region of the parameter space near the transition 
to the normal state \cite{FFLO} and usually is destroyed when the exchange 
splitting $E_{ex}$ is larger than $\sqrt{2}/2 \Delta$ (Clogston criterion) 
\cite{Clogston}, where $\Delta$ is the $SC$ energy gap. Moreover the $FFLO$ 
state is very sensitive to both elastic and spin-orbit scattering 
\cite{Takada}. The last two effects make the $FFLO$ state very difficult to 
observe experimentally in a bulk samples. The situation is much more favorable 
in $FM$/$SC$ heterostructures where, due to the proximity effect, the Cooper 
pairs can survive even if the exchange field in $FM$ is much larger than $SC$ 
gap.

According to our current understanding of the $FFLO$ phenomenon in a $FM$/$SC$
structure, when a Cooper pair enters the ferromagnet it acquires a center of
mass momentum $\hbar Q = 2 E_{ex} / v_F$ \cite{Demler}, where $v_F$ is Fermi 
velocity. Usually the $SC$ phase changes linearly with distance $x$ from the 
interface, $\phi = Q_x x$, and this results in a oscillatory behavior of the 
pairing amplitude in the $FM$. It turns out that under certain conditions a 
$3D$-$FFLO$ state, featuring a spatial dependence of the pairing amplitude also 
along the interface, can be realized  \cite{Izyumov_1,KGA}. It is this latter 
case that we shall deal with here. 

The purpose of the present paper is to demonstrate that in a ferromagnetic 
layer on a superconducting substrate, for particular values of the exchange
splitting $E_{ex}$ and the layer thickness $d_F$, a $FFLO$ like state is
realized with the pairing amplitude $f$ varying both perpendicular and parallel
to the $FM$/$SC$ interface. It will be shown that such a ground state supports 
a spontaneously generated current flowing in opposite directions in the $FM$ 
and the $SC$ regions. The existence of this remarkable state was first 
predicted on the basis of a simple lattice model \cite{KGA}. Here we shall deal 
with the problem by a less model dependent, semiclassical approach and address 
the issue of the observability of the phenomenon in the presence of disorder 
within $FM$ layer and at the $FM$/$SC$ interface. 

\begin{figure}[h]
 \resizebox{8.4cm}{5.5cm}{
  \includegraphics{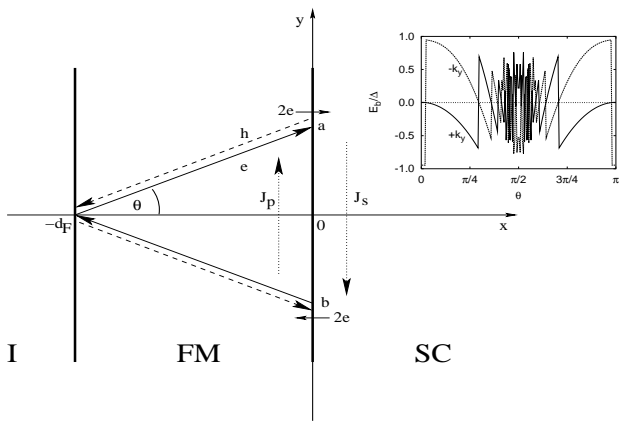}}
 \caption{\label{Fig1} The (Bohr-Sommerfeld) semiclassical trajectories for 
          quasiparticles which scatter spectacularly at the $I$/$FM$ interface 
	  and by Andreev reflections at the $FM$/$SC$ interface. Note that the 
	  trajectories have particle-like and hole-like segments and they imply 
	  an electric current, $J_p$ in the $y$-direction. Inset: The energy of 
	  the Andreev bound states associated with spin up electrons moving in
	  positive ($+k_y$) and negative ($-k_y$) $y$-direction as a function 
	  of $\theta$ for $\xi_F/d_F = 0.425$ and $\phi=0$. Evidently, some of 
	  these bound states are at zero energy.}
\end{figure}

The system we consider is sketched in Fig \ref{Fig1}. It consists of thin 
ferromagnet ($FM$) of thickness $d_F$ deposited on a semi-infinite 
superconductor ($SC$) and bounded on the other side by an insulator.
In such an $I$/$FM$/$SC$ quantum well there will be bound states corresponding 
to the closed quasiparticle trajectories \cite{deGennes}. Each trajectory 
consists of an electron segment, $e$, which includes an Andreev reflection at 
the $FM$/$SC$ interface and an ordinary reflection at the $I$/$FM$ interface 
plus a hole segment, $h$, retracing back the electron trajectory 
(see Fig \ref{Fig1}). The Bohr-Sommerfeld quantization rule \cite{Kevin}:
\begin{eqnarray}
 \int^a_b p_e(\omega) dl - \int^b_a p_h(\omega) dl + \delta \phi - 
 \gamma(\omega) = 2 n \pi
 \label{BS}
\end{eqnarray}
gives the energies of the bound states. The first and second terms represent 
the total phase accumulated during propagation through the $FM$ region from $b$ 
to $a$ ($=4 d_F (\omega + \sigma E_{ex})/v_F cos{(\theta)}$), $\delta \phi$ is 
the phase difference between points $b$ and $a$, and 
$\gamma(\omega) = arccos(\omega/\Delta)$ is the Andreev reflection phase shift.
An example of the $\theta$-dependence of the Andreev bound state ($ABS$) 
energies for $\xi_F/d_F=0.425$, where $\xi_F=\hbar v_F/E_{ex}$ is the $FM$
coherence length, and $\delta \phi = 0$ is shown in the inset of Fig. 
\ref{Fig1}. Clearly, for any exchange splitting $E_{ex} \neq 0$ it is possible 
to find such a $\theta$ that the corresponding $ABS$ is exactly at zero energy. 
If there is a large number of such zero-energy $ABS$ for some exchange 
splittings (more precisely for $\xi_F/d_F$ ratio) the density of states $DOS$ 
has a large peak at the Fermi energy ($\epsilon_F = 0$). Such a situation turns 
out to be energetically unfavorable. There is a number of mechanisms which 
split this peak and thereby lower the energy of the system \cite{Sigrist}. One 
of these is a spontaneous current which 'Doppler' shifts the quasiparticle 
energies by $\delta = e v_F A_y cos(\theta)$ \cite{Higashitani}, where $A_y$ is 
a vector potential in the $y$ direction.

\begin{figure}[h]
 \resizebox{8.4cm}{5.5cm}{
  \includegraphics{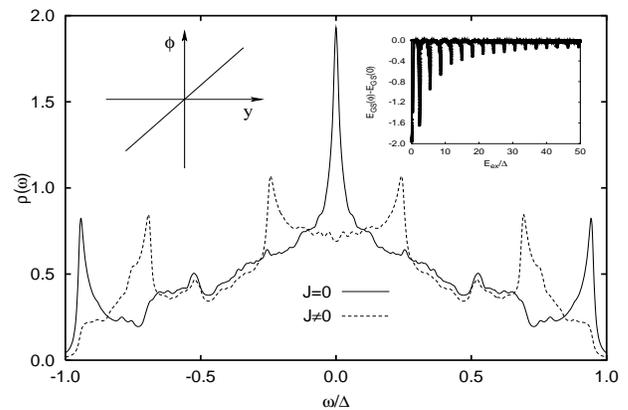}}
 \caption{\label{Fig2} The total density of states $\rho(\omega)$ for 
          $\xi_F/d_F = 0.425$ with no current flow (solid line) and in the 
	  presence of the current (dashed line). For 
	  $J \propto \partial \phi/\partial y \neq 0$ the zero energy state 
	  splits and this lowers the energy of the system. Right inset: The 
	  difference between total energies of the $ABS$ with current flow 
	  $E_{GS}(\phi)$ and without the current $E_{GS}(0)$ as a function of 
	  $E_{ex}/\Delta$ ($d_F/\xi_S=1$). Left inset: The phase of the 
	  superconducting order parameter $\phi$ as a function of $y$. The 
	  slope $\partial \phi/\partial y$ implies a supercurrent $J_s$ which 
	  is carried by Cooper pairs from point $a$ to $b$ 
	  (see Fig. \ref{Fig1}).}
\end{figure}

An example of such a density of states is depicted in Fig. \ref{Fig2}, where 
one can see large peak at zero energy (solid line). 
This corresponds to zero $SC$ phase difference between points $a$ and $b$ in
Fig. \ref{Fig1}, namely no spontaneous current. In the inset of Fig. \ref{Fig2} 
the energy difference between state with spontaneous current $E_{GS}(\phi)$ and 
state with no current $E_{GS}(0)$ is shown. The corresponding $DOS$, with the 
zero energy $ABS$ split, is plotted with the dashed line in Fig. \ref{Fig2}. 

To summarize, we have shown, using a Bohr-Sommerfeld semiclassical argument, 
that for any exchange splitting, $E_{ex}$, there are Andreev bound states at 
zero energy and for certain $E_{ex}$ the number of such states is so large that 
it produces huge zero energy peak in the density of states. Such a peak in turn 
is split by a spontaneous current and this lowers the total energy of the 
system. This current carrying state can be regarded as a realization of the 
$FFLO$ variation of the pairing amplitude in the $y$-direction. So, one can say 
that the system can be switched between $1D$ and $2D$ $FFLO$-like states as the 
exchange field or thickness of the ferromagnet is changed. In the following we 
will show that this spontaneous current can be also obtained within a 
self-consistent quasiclassical theory. This approach allows for a treatment of 
disorder and finite transparency of the interface, and hence provides further 
useful insights from the experimental point of view. 

The quasiclassical matrix Eilenberger equation \cite{Eilenberger} reads
\begin{eqnarray}
 {\bf v}_F {\bf \nabla} \hat g_{\sigma}({\bf v}_F,{\bf r}) + 
 \nonumber \\
 \left[\tilde \omega_{\sigma}({\bf r}) \hat \tau_3 + \hat \Delta({\bf r}) 
     + \frac{1}{2\tau} \langle \hat g_{\sigma}({\bf v}_F,{\bf r}) \rangle,
       \hat g_{\sigma}({\bf v}_F,{\bf r}) \right]_- = 0
 \label{Eilenb}
\end{eqnarray}
where 
\begin{eqnarray}
 \hat g_{\sigma} = \left(
 \begin{array}{lr}
  g_{\sigma} & f_{\sigma} \\
  f^+_{\sigma} & -g_{\sigma}
 \end{array}
 \right), \;\;\;\;\; 
 \hat \Delta = \left(
 \begin{array}{lr}
  0 & \Delta \\
  \Delta^* & 0
 \end{array}
 \right) \;\;\; 
 \label{g}
\end{eqnarray}
and 
\begin{eqnarray}
 \tilde \omega_{\sigma}({\bf r}) = \omega + i \sigma E_{ex} + 
  i e {\bf v}_F {\bf A}({\bf r})
 \label{omega}
\end{eqnarray}
Here $\hat \tau_3$ is the Pauli matrix, $\omega = \pi T (2 n + 1)$ is the
Matsubara frequency, $\sigma$ ($= \pm 1$) labels the electron spin, 
${\bf A}({\bf r})$ is the vector potential and $\langle ... \rangle$ denotes
averaging over directions of the Fermi velocity ${\bf v}_F$. The matrix 
Green's function has to obey the normalization condition 
$\hat g^2_{\sigma}({\bf v}_F,{\bf r}) = \hat 1$. The exchange splitting 
$E_{ex}$ is non-zero and constant in the ferromagnet only while 
$\Delta({\bf r})$ is non-zero in the superconductor and is calculated 
self-consistently from 
\begin{eqnarray}
 \Delta({\bf r}) = U \pi \rho_0 T \sum_{\omega,\sigma}
 \langle f_{\sigma}({\bf v}_F,{\bf r}) \rangle,
 \label{Delta}
\end{eqnarray}
where we have assumed that the coupling constant $U < 0$ in the $SC$ and $= 0$ 
in the $FM$. $\rho_0$ is the normal state $DOS$ and $T$ stands for temperature. 

The spontaneous current has to be determined self-consistently together with 
the Maxwell equation (Ampere's law), which couples the electron current to the 
magnetic field. The total current in the $y$-direction at each point $x$ 
measured from the interface is given by 
\begin{eqnarray}
 J^{tot}_y(x) = 2 i e \pi \rho_0 T \sum_{\omega,\sigma}
 \langle {\bf v}_F g_{\sigma}({\bf v}_F,x) \rangle, 
 \label{current}
\end{eqnarray}
where $e$ is the electron charge, while the Maxwell equation (in the Landau
gauge) reads
\begin{eqnarray}
 \frac{d^2 A_y(x)}{d x^2} = - \mu_0 J^{tot}_y(x)
 \label{Maxwell}
\end{eqnarray}
with $\mu_0$ being the permeability of free space. 

We have solved the Eilenberger equation (\ref{Eilenb}) numerically along each 
quasiparticle $2D$ trajectory using the Riccati parametrization (Schopohl-Maki 
transformation) \cite{Riccati} together with the self-consistency relations 
(\ref{Delta})-(\ref{Maxwell}).

\begin{figure}[h]
 \resizebox{8.4cm}{5.5cm}{
  \includegraphics{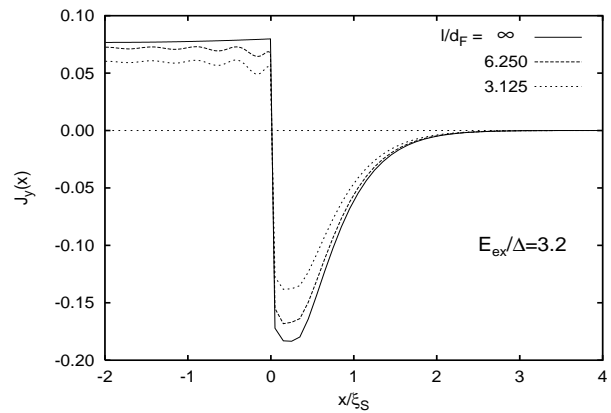}}
 \caption{\label{Fig3} The total spontaneous current as a function of $x$ for 
          a number of mean free path values $l$. Note that disorder introduces
	  oscillations of the current.}
\end{figure}

The most remarkable feature of the self-consistent solution is that the
iterations of the Eilenberger equations frequently converge to a solution with 
a finite value of the current even though there is no external vector 
potential. The current flows in one direction over the whole ferromagnet and 
flows back on $SC$ side on the scale of $SC$ coherence length $\xi_S$ (see 
solid curve in the Fig.\ref{Fig3}), so the total current is zero, as it should 
be in the true ground state. Such a distribution of the current can be 
associated with a quasiparticle current in $FM$ and with a supercurrent on the 
$SC$ side mediated by Cooper pairs (see Fig. \ref{Fig1}). The fact that the 
current flows over the whole $FM$ is due to the extended nature of the $ABS$. 
There is also magnetic flux associated with such current distribution. 
Typically the spontaneous magnetic field produced by this current is of order 
of $0.1 \; B_{c2}$, where $B_{c2}$ is the upper critical field of the bulk 
$SC$. 

Within the present self-consistent calculations we were also able to study the
effect of elastic scattering in the $FM$. The disorder in the ($s$-wave) $SC$ 
can be neglected due to Andreson's theorem \cite{Moradian}. The current for a 
number of mean free paths is shown in Fig. \ref{Fig3}.
As we one can see disorder introduces oscillations of the current. The
spontaneous current is proportional to the $DOS$ at the Fermi energy \cite{KGA}
and so the oscillations of the current are related to the oscillatory behavior 
of $DOS$ in the disordered sample \cite{Zareyan,Buzdin_1}. In the clean limit 
the $DOS$ is constant in the whole $FM$. As is well known, this is a property 
of the Eilenberger equations in the clean limit \cite{Buzdin_1}. Moreover 
disorder also suppresses the current, as expected, since it introduces 
decoherence of electron-hole pairs in the $FM$. Finally if the mean free path 
$l$ is shorter than the $FM$ thickness $d_F$ the current is completely 
suppressed. If $l < d_F$ the Andreev reflected particles cannot reach the 
$I$/$FM$ interface, which is a necessary condition for the formation of the 
current carrying $ABS$, because they are scattered on the impurities and 
the electron-hole coherence is lost. In this regard the $FFLO$ variation of the 
pairing amplitude in the $y$-direction is very sensitive to the elastic 
disorder. However in the $x$-direction $FFLO$ state persists until $l < \xi_F$ 
\cite{Demler}, even if $l < d_F$. 

\begin{figure}[h]
 \resizebox{8.4cm}{5.5cm}{
  \includegraphics{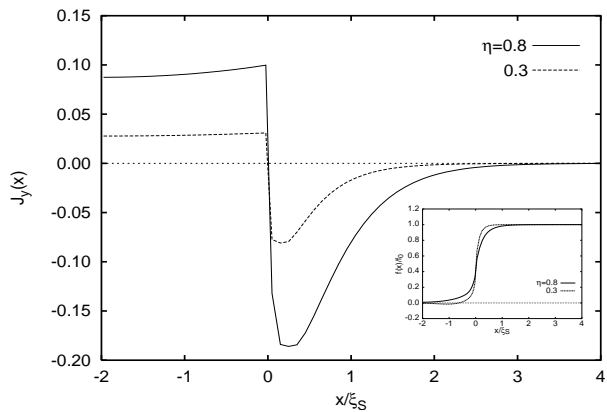}}
 \caption{\label{Fig4} Effect of finite transparency $\eta$ on the spontaneous 
          current $E_{ex}/\Delta = 0.2$. $\eta = 1$ means a reflectionless
	  interface, while $\eta = 0$ a perfectly reflective one. Inset: The 
	  corresponding pairing amplitude normalized to its bulk value. Note 
	  that $\eta$ changes the period of oscillations of $f(x)$.}
\end{figure}

To take into account the effect of specular reflections at the $FM$/$SC$ 
interface we adapt the approach proposed by M. Zareyan {\it et al.} 
\cite{Zareyan}, where a certain probability distribution was associated with 
each semiclassical trajectory (for details see Ref. \cite{Zareyan}). In Fig.
\ref{Fig4} we show the current for two different transparencies $0 < \eta < 1$. 
As we would expect transparency $\eta < 1$ suppressed the current because it
suppresses Andreev reflection processes and at the same time introduces normal
(specular) reflections at the $FM$/$SC$ interface. But even for $\eta = 0.3$, 
in the present case, we still get the current. It turns out that $\eta$ has
similar influence on the properties of the system as $E_{ex}$ does. As we can
read from the inset of Fig. \ref{Fig4}, it changes the period of oscillations 
of the pairing amplitude. Moreover if we changed $\eta$ only, for certain
its values we get solutions with a current flowing, whilst for others 
solutions with no current. For the above set of parameters we have found that 
current flows in the regions where $0.7 < \eta < 1$ and $0.2 < \eta < 0.35$. So 
we can also switch between the $1D$ and $2D$ $FFLO$ state by changing the 
transparency. 

Finally we note that so far we took no account of the demagnetization field
$B_d$ due to the ferromagnet. In the limit of a zero film width this is zero. 
For finite thickness, $d_F$, and magnetization, $M$, (which in the Stoner model
\cite{Mohn} corresponds to $E_{ex} = I M$, where $I$ is a fenomenological
parameter) we estimate $B_d$ to be considerably less then $B_{sp}$ due to the
spontaneous current. Furthemore it should be stressed that in all our 
calculations the magnetization was constrained to point in the $z$ direction. 
Thus the direction of the spontaneous current was determined by the condition 
that $B_{sp}$ is parallel to $M$. In a more general theory where Andreev orbits 
also occur in the $x$-$z$ plane and spin orbit coupling is taken into account 
these issues would need to be reexamined. Also the above calculations were two
dimensional, but preliminary studies of a $3D$ system indicate that there are 
no qualitative changes when orbits in the $x$-$z$ plane are included. 

In summary we have demonstrated that under certain conditions the ground state
of a $I$/$FM$/$SC$ trilayer features a spontaneous current flowing in opposite
directions in the $FM$ and $SC$ layers. We argued that this state can be viewed 
as a $2D$ $FFLO$ proximity state and hence the observation of the above current
would be a decisive proof that the surprising behavior of such heterostructures
is governed by the $FFLO$ phenomenon. We also showed that this state persists 
only in the clean limit where the mean free path is longer than $FM$ thickness 
and investigated the effect of low transparency of the interface on the 
observibility of the ground state current.

One of us, M.K., would like to thank Prof. Yuli Nazarov for helpful discussion
on the problem of transparency of the interface. 
This work has been supported by Computational Magnetoelectronics Research
Training Network under Contract No. HPRN-CT-2000-00143.



\begin{thebibliography}{99}
%
\bibitem{Chien} C. L. Chien, D. H. Reich, J. Magn. Magn. Mater. {\bf 200}, 83
                (1999); 
		Y. A. Izyumov {\it et al.}, Phys. Usp. {\bf 45}, 109 (2002).
%
\bibitem{Lambert} C. J. Lambert, R. Raimondi, J. Phys. Condens. Matter
                  {\bf 10}, 901 (1998);
                  C. W. J. Beenakker, Rev. Mod. Phys. {\bf 69}, 731 (1997);
                  B. Pannetier, H. Courtois, J. Low Temp. Phys. {\bf 118}, 599
                  (2000).
%
\bibitem{Andreev} A. F. Andreev, Sov. Phys. JETP {\bf 19}, 1228 (1964).
%
\bibitem{Berk} N. F. Berk, J. R. Schrieffer, Phys. Rev. Lett. {\bf 17}, 433
               (1966); 
	       C. Pfleiderer {\it et al.}, Nature {\bf 412}, 58 (2001);
	       D. Aoki {\it et al.}, Nature {\bf 413}, 613 (2001).
%
\bibitem{Bauer} G. E. W. Bauer {\it et al.}, Materials Sci. Eng. {\bf B84}, 31
                (2001); 
		S. Oh {\it et al.}, Appl. Phys. Lett. {\bf 71}, 2376 (1997); 
		L. R. Tagirov, Phys. Rev. Lett. {\bf 83}, 2058 (1999);
		Y. N. Proshin {\it et al.}, Supercond. Sci. Technol. {\bf 15},
		285 (2002).
%
\bibitem{Blatter} G. Blatter {\it et al.}, Phys. Rev. {\bf B63}, 174511 (2001).
%
\bibitem{Wong} H. K. Wong {\it et al.}, J. Low Temp. Phys. {\bf 63}, 307 
               (1986); 
	       J. S. Jiang {\it et al.}, Phys. Rev. Lett. {\bf 74}, 314 (1995); 
	       V. Marcaldo {\it et al.}, Phys. Rev. {\bf 53}, 14 040 (1996); 
	       Th. M\"{u}hge {\it et al.}, Phys. Rev. Lett. {\bf 77}, 1857 
	       (1996).
%
\bibitem{Buzdin} A. I. Buzdin {\it et al.}, JETP Lett. {\bf 35}, 178 (1982); 
                 A. I. Buzdin, M. V. Kuprianov, JETP Lett. {\bf 52}, 487 (1990).
%
\bibitem{Radovic} Z. Radovi\'{c} {\it et al.}, Phys. Rev. {\bf B44}, 759 (1991).
%
\bibitem{Demler} E. A. Demler {\it et al.}, Phys. Rev. {\bf B55}, 15 174 (1997).
%
\bibitem{Kontos} T. Kontos {\it et al.}, Phys. Rev. Lett. {\bf 86}, 304 (2001).
%
\bibitem{Zareyan} M. Zareyan {\it et al.}, Phys. Rev. Lett. {\bf 86}, 308 
                  (2001).
%
\bibitem{Buzdin_1} A. I. Buzdin, Phys. Rev. {\bf B62}, 11 377;
                   I. Baladi\'{e}, A. I. Buzdin, Phys. Rev. {\bf B64}, 224514
                   (2001); 
		   K. Halterman, O. T. Valls, Phys. Rev. {\bf 65} 014509 (2002);
		   F. S. Bergeret {\it et al.}, Phys. Rev. {\bf B65}, 134505 
		   (2002).
%
\bibitem{KGA} M. Krawiec {\it et al.}, Phys. Rev. {\bf B66}, 172505 (2002); 
              preprint cond-mat/0207135.
%
\bibitem{Lee} S. F. Lee {\it et al.}, preprint cond-mat/0301313.
%
\bibitem{Volkov} A. F. Volkov {\it et al.}, preprint cond-mat/0212384; 
                 M. Eschrig {\it et al.}, preprint cond-mat/0206278.
%
\bibitem{FFLO} P. Fulde, A. Ferrell, Phys. Rev. {\bf 135}, A550 (1964); 
               A. Larkin, Y. Ovchinnikov, Sov. Phys. JETP {\bf 20}, 762 (1965).
%
\bibitem{Clogston} M. A. Clogston, Phys. Rev. Lett. {\bf 9}, 266 (1962).
%
\bibitem{Takada} S. Takada, Prog. Theor. Phys. {\bf 43}, 27 (1970)
%
\bibitem{Izyumov_1} Y. A. Izyumov {\it et al.}, JETP Lett. {\bf 71}, 138 
                    (2000);
                    M. G. Khusainov {\it et al.}, Physica {\bf B284-288}, 503 
		    (2000).
%
\bibitem{deGennes} P. G. de Gennes and D. Saint-James, Phys. Lett. {\bf 4}, 151
                   (1963).
%
\bibitem{Kevin} K. P. Duncan, B. L. Gy\"{o}rffy, Ann. Phys. (New York) 
                {\bf 298}, 273 (2002).
%
\bibitem{Sigrist} M. Sigrist, Prog. Theor. Phys. {\bf 99}, 899 (1998); 
                  M. Fogelstr\"{o}m, S. -K. Yip, Phys. Rev. {\bf B57},
                  R14 060 (1998)
%
\bibitem{Higashitani} S. Higashitani, J. Phys. Soc. Jpn. {\bf 66}, 2556 (1997).
%
\bibitem{Eilenberger} G. Eilenberger, Z. Phys. {\bf 214}, 195 (1968).
%
\bibitem{Riccati} N. Schopohl, K. Maki, Phys. Rev. {\bf B52}, 490 (1995); 
                  N. Schopohl, cond-mat/9804064.
%
\bibitem{Moradian} R. Moradian {\it et al.}, Phys. rev. {\bf B63}, 024501 
                   (2001).

\bibitem{Mohn} P. Mohn, {\it Magnetism in the Solid State}, Springer-Verlag,
               Berlin and Heidelberg 2003.
\end{thebibliography}
\end{document}